# Magnetic Domains and Surface Effects in Hollow Maghemite Nanoparticles

Andreu Cabot and A. Paul Alivisatos*

*Department of Chemistry, University of California at Berkeley, and Materials Sciences Division, Lawrence Berkeley National Laboratory, Berkeley, CA 94720*

Víctor F. Puntes

*Institut Català d' Estudis i Recerca Avançat and Institut Catalá de Nanotecnologia E-08193 Bellaterra. Barcelona*

Lluís Balcells

*Institut de Ciència de Materials de Barcelona, CSIC, Campus UAB, Bellaterra-08193, Spain*

Óscar Iglesias and Amílcar Labarta

*Departament de Física Fonamental and Institut de Nanociència i Nanotecnologia, Universitat de Barcelona, Martí i Franquès 1, 08028 Barcelona, Spain*

## Abstract

In the present work, we investigate the magnetic properties of ferrimagnetic and non-interacting maghemite ($\gamma$-$Fe_2O_3$) hollow nanoparticles obtained by the Kirkendall effect. From the experimental characterization of their magnetic behavior, we find that polycrystalline hollow maghemite nanoparticles exhibit low blocked-to-superparamagnetic transition temperatures, small magnetic moments, significant coercivities and irreversibility fields, and no magnetic saturation on external magnetic fields up to 5 T. These results are interpreted in terms of the microstructural parameters characterizing the maghemite shells by means of atomistic Monte Carlo simulations of an individual spherical shell. The model comprises strongly interacting crystallographic domains arranged in a spherical shell with random orientations and anisotropy axis. The Monte Carlo simulation allows discernment between the influence of the polycrystalline structure and its hollow geometry, while revealing the magnetic domain arrangement in the different temperature regimes.

*Corresponding author. E-mail:* * alivis@berkeley.edu

## I. INTRODUCTION

In extended materials, the strength and length scale of typical spin-spin interactions are such that ordering of spins frequently occurs over ranges with sizes in the nanometer scale. In nanoparticles, however, the crystal size and geometry determine the extent and configuration of the magnetic domains. In polycrystalline nanostructures and nanoparticle arrays, the competition between the crystallographic anisotropy and the strength of the spin-spin interaction between neighboring crystals, determines the magnetic behavior of the composites. This competition relies not only on the size and shape of the crystallographic domains, but also on their relative orientation and geometric organization.

Due to such dependencies of the magnetic properties, advances in the ability to pattern matter on the nanometer scale have created new opportunities to develop magnetic materials with novel characteristics and applications.[1-4] One such novel type of magnetic material design, which has recently attracted significant attention,[5-8] is the hollow geometry. D. Goll et al. showed that the hollow geometry incorporates additional parameters for the tuning of the magnetic properties of the nanoparticles.[6] They theoretically determined the phase diagram of the lowest-energy domain configurations in hollow ferromagnetic nanoparticles as a function of the material parameters, particle size and the shell thickness.[6] However, while this initial model did not include interface or surface effects, actual hollow nanoparticles are characterized by large surface to bulk ratios. Moreover, hollow nanoparticles synthesized by the Kirkendall effect[8-11] or by means of templates,[12-14] are usually polycrystalline structures, due to the multiplicity of shell nucleation sites. Thus, they have multiple crystallographic domains, which are randomly oriented and so have differentiated local anisotropy axes.

In the present work, we study the magnetic properties of polycrystalline hollow maghemite nanoparticles obtained by the Kirkendall effect. We experimentally analyze their magnetic behavior and interpret our experimental results using an atomistic Monte Carlo simulation of a model for an individual maghemite nanoshell.

## II. HOLLOW NANOPARTICLES AND STRUCTURAL CHARACTERIZATION

Hollow maghemite nanoparticles were obtained following a previously reported procedure based on the Kirkendall effect.[8] Briefly, iron pentacarbonyl was decomposed in air-free conditions at around 220ºC in organic solvents containing surfactants. The resulting iron-based nanoparticles were oxidized in solution by means of a dry synthetic air flow. Owing to the faster self-diffusion of iron than oxygen ions within iron oxide, the oxidation of 1-20 nm iron nanoparticles results in hollow iron oxide nanostructures.

Hollow iron oxide nanoparticles obtained by the Kirkendall effect have an inner-to-outer diameter ratio of around $\phi_I/\phi_E = 0.6$ and relatively narrow particle size distributions. The hollow nanoparticles studied in this work have a diameter of 8.1±0.6 nm with 1.6±0.2 nm thick shells and a size dispersion of around 10%. Figure 1(a) shows a transmission electron micrograph of the hollow iron oxide nanoparticles supported on a carbon grid. Further high resolution TEM characterization of the particles show them to be crystalline, but to contain multiple crystallographic domains within each shell (Fig. 1(b) and 1(c)). Each hollow nanoparticle is composed of approximately 10 crystallographic domains having random orientations. The presence of intergrains in the shell may allow for surfactants and solvent to enter inside the particle, thus there may not be a true void inside these structures, but it may be filled with organic solvents in solution and with gas in air. The crystallographic structure of the hollow nanoparticles was identified as that of maghemite by X-ray absorption spectroscopy.[8]

## III. MAGNETIC PROPERTIES

Prior to magnetic characterization, the solution containing the maghemite shells was centrifuged to remove any possible particle aggregates. For magnetic characterization, maghemite nanoshells were dispersed in a 50% mixture of high melting point organic solvents, namely: nonadecane ($C_{19}H_{40}$, $T_m$ = 32 ºC) and dotriacontane ($C_{32}H_{66}$, $T_m$ = 69 ºC). In order to avoid interparticle interactions, the particle concentration was kept at about 0.2-0.3 % in mass, as measured by means of ionic-coupled mass spectroscopy. The magnetic measurements were carried out in an XL Quantum Design superconducting quantum interference device (SQUID) using 0.2 g of the diluted sample.

Figure 2 shows the magnetic susceptibility vs. temperature for the maghemite nanoshells following zero-field-cool (ZFC) and field-cool (FC) processes. The close coincidence of the ZFC peak and the onset of the irreversibility between the ZFC and FC magnetization curves allow us to exclude a large extent of particle aggregation or large size distributions, which is consistent with the TEM characterization of the sample (see Fig. 1). For the low concentration range used in our experiments, the temperature at the ZFC peak is about 34 K and independent of the particle concentration, which excludes interparticle interactions.[15] This value of the temperature of the ZFC peak is lower than the blocking temperature observed in 7 nm solid maghemite particles, which have a particle volume, and thus a number of spins, equivalent to that of the 8.1 nm hollow particles (roughly 200 $nm^3$ and $8x10^3$ Fe atoms per particle).[16] However, this value of the temperature of the ZFC peak is larger than that corresponding to isolated maghemite crystallites of about 21 $nm^3$ (about 3.4 nm in diameter assuming spherical shape), equivalent in size to those forming the shell (inset to Figure 2). This experimental observation indicates that either (a) magnetic interactions among crystallites within each hollow particle yield magnetic frustration, which increases the effective blocking temperature of the crystallite, or (b) there is an enhanced value of the anisotropy energy per unit volume with respect to that of solid nanoparticles with similar magnetic volumes.

The study of the particle magnetization as a function of the observational time window, by means of ac susceptibility measurements, is a conventional method to evaluate the average magnetic anisotropy barrier per particle (Fig. 3). For a given measuring frequency ($\nu$) and particle size distribution, the real part of the ac susceptibility ($\chi$') peaks at a temperature ($T_{max}$) such, that the measuring time ($\tau=1/\nu$) coincides with the relaxation time of those magnetic domains having the average anisotropy energy and size. Taking into account that $T_{max}$ and the attempt time are related through the Arrhenius' law, the mean value of the anisotropy energy can be evaluated by linear regression of $\tau$ as a function of $1/T_{max}$ (see inset to Fig. 3). For the hollow particles, this regression yields an anisotropy energy per unit volume of $7x10^6$ $erg/cm^3$. Such a magnetic anisotropy constant is one order of magnitude larger than that of solid nanoparticles with a similar number of spins (7 nm in diameter assuming spherical shape),[17] and two orders of magnitude larger than that of bulk maghemite ($4.7x10^4$ $erg/cm^3$).[18] It is commonly agreed that, at the surface, the broken translational symmetry of the crystal and the lower coordination leads to a stronger anisotropy than in the bulk. Anisotropy energies per atom at the surface are usually two or three orders of magnitude larger than in bulk materials, yielding an anisotropy enhancement in nanoparticles and thin films.[19-21]. Thus, we associate the huge particle anisotropy obtained for hollow maghemite nanoparticles to the large proportion of spins with lower coordination, located at the innermost or outermost surfaces of the shell and at the interfaces between crystallographic domains.

Figure 4(a) shows the hysteresis loop of hollow particles at 5 K. It evidences that hollow particles are characterized by high values of the coercive field and the irreversibility field (the field at which the decreasing and increasing field loop branches join). The coercive field is around 3300 Oe and the irreversibility field is larger than the maximum applied field (50 kOe). In fact, the hysteresis loop in Fig. 4(a) resembles those of frustrated and disordered magnets, such as random anisotropy systems. We attribute this behavior to the polycrystalline nature of the maghemite shells and the large number of spins pinned by surface anisotropy effects. At low temperatures, spins tend to align parallel to the crystalline anisotropy axes existing in each individual crystallite. Such a tendency leads to the formation of multiple magnetic domains within each shell, instead of a single domain with all the spins aligned along a unique axis as was predicted by D. Goll et al. for single crystal nanoshells of diameter below 10-20 nm.[6] Besides, there also exists a significant high-field linear contribution to the magnetization, arising from the spins at the shell surface and crystallite interfaces, which are strongly pinned along local axes due to surface anisotropy. The saturation magnetization associated with the spins at the crystallite cores, which are those remaining with ferrimagnetic ordering like in bulk maghemite, can be estimated to be about 3-4 emu/g by linear extrapolation to zero field of the hysteresis loop at high fields (Fig. 4(a)). This value is about 20 times smaller than that corresponding to the bulk counterpart (74 emu/g), what gives a clear indication of the high magnetic frustration and high fraction of surface spins present in the hollow particles. Such magnetic frustration, arising from the existence of magnetic domains and surface anisotropy effects, is at the origin of the observed high irreversibility and coercive field of the polycrystalline hollow nanoparticles. In addition, a strong shift of the hysteresis loop, over 3000 Oe, is observed when cooling the particles in the presence of a magnetic field. Note that, in these experiments, the maximum applied field is lower than the irreversibility field, so the observed loop shift may not correspond to an exchange bias phenomenon, but just to a minor loop of the hysteresis loop.

The saturation magnetization of the ferrimagnetic component of the hollow nanoparticles at low temperatures is significantly lower than that observed in solid nanoparticles of similar size or in bulk maghemite. We can gain further insight in this reduced value of the saturation magnetization by analyzing the magnetization curves in the superparamagnetic (SPM) regime. In the SPM regime, the crystal anisotropy barriers of the crystallites composing each nanoshell are overcome by thermal excitation. In this scenario, it is expected that core spins of all the crystallites in each shell magnetize as a whole following the external applied field. Therefore, we can estimate the mean value of the ferrimagnetic component of the hollow particles' magnetization (corresponding to the cores of the crystallites) by fitting a log-normal distribution of Langevin functions L(x) plus a paramagnetic contribution to the magnetization vs. field curve at 200 K (when the sample is clearly in the SPM regime):

$$M(H,T) = \int dm\, m P(m) L(mH / k_B T) + \chi_d H \qquad (1)$$

where m is the magnetic moment per particle and $\chi_p$ is a paramagnetic susceptibility (Fig. 4(b)). The obtained distribution of magnetic moments P(m) is shown in the inset to Fig. 4(b). The mean magnetic moment per hollow particle of this distribution is $3.3 \times 10^{-18}$ emu (360 $\mu_B$, where $\mu_B$ is the Bohr magneton). This magnetic moment is equivalent to 9 $nm^3$ of bulk maghemite (74 emu/g), which is a volume 24 times smaller than that of the total material volume per hollow nanoparticle. It is worth noting that the saturation magnetization of the ferrimagnetic component deduced from this fitting is about 3 emu/g, which is in good agreement with the value estimated from the hysteresis loop. From the fitting of the magnetization curve at 200 K, a large paramagnetic susceptibility $\chi_p$ is also obtained (see linear contribution in Fig. 4(b)). This very large high-field susceptibility is consistent with the shape of the hysteresis loop at 5 K.

The very low saturation magnetization and the high paramagnetic susceptibility are explained by the large disorder on the hollow nanoparticles ubiquitous surface and crystallographic interfaces, which leads to the reduction of the number of spins aligning with the external field.[22,23] Furthermore, aside from the spin disorder at the nanoparticles surface, the ferrimagnetic character of maghemite has associated significant finite size effects:[24-26] Maghemite's net magnetic moment arises from the unbalanced number of spins in an antiparallel arrangement. In the nanoscale, this balance can differ from that of the bulk material, leading to a significant reduction of the magnetization.

From an experimental point of view, the shell magnetization can be increased by improving the shell crystalline structure in two ways: i) An increase of the synthesis temperatures or a-posteriori sintering process would lead to less defective and larger crystallographic domains. However, the growth of the crystallographic domains within the shell is limited by the shell thickness and thus by the particle size. An excessive growth of the crystallographic domains within the shell leads to its rupture.[8] ii) Larger hollow particles, having a thicker shell, would provide larger crystal domain sizes, while at the same time allowing synthesis or sintering treatments at higher temperatures, thus reaching better crystallinity. However, the size of the maghemite hollow particles obtained by the Kirkendall effect is limited by the iron diffusion inside the shell, as previously reported.[8]

## IV. MONTE CARLO SIMULATION

In order to elucidate the origin of the magnetic characteristics of the hollow maghemite nanoparticles, we have carried out atomistic Monte Carlo simulations of an individual maghemite nanoshell model. In our model, the magnetic ions are represented by classical Heisenberg spins placed on the nodes of the real maghemite structure sublattices, having tetrahedral and octahedral coordinations and interacting according to the following Hamiltonian:

$$H / k_B = -\sum_{\langle i,j \rangle} J_{ij} \left( \vec{S}_i \cdot \vec{S}_j \right) - \sum_i \vec{h} \cdot \vec{S}_i + E_{anis} \qquad (2)$$

The first term is the nn exchange interaction, the second is the Zeeman energy with $h = \mu H/k_B$ (H is the magnetic field and μ the magnetic moment of the magnetic ion), and the third corresponds to the magnetocrystalline anisotropy energy.[27] In this last term, we have distinguished surface spins, having reduced coordination with respect to bulk and anisotropy constant $k_S$, from the core spins, having full coordination and an anisotropy constant $k_C$. We consider a Neél type anisotropy for the surface spins and a uniaxial anisotropy along the direction $\hat{n}_i$ for the core spins. The corresponding energy can be expressed as:

$$E_{anis} = k_S \sum_{i \in S} \sum_{j \in nn} \left( \vec{S}_i \cdot \hat{r}_{ij} \right)^2 - k_C \sum_{i \in C} \left( \vec{S}_i \cdot \hat{n}_i \right)^2 , \qquad (3)$$

where $\hat{r}_{ij}$ is a unit vector joining spin *i* with its nearest neighbors *j* and $\hat{n}_i$ is the anisotropy axis of each crystallite. The simulated hollow spherical nanoparticles have a total radius of 4.88 *a* (where *a* is the cell parameter of the maghemite) and a shell with thickness $D_{Sh}$ varying between 1.92 *a* (actual thickness of the hollow particles experimentally studied in this work) and 4.88 *a* (filled particle). In order to better model the structure of the real particles, the spherical nanoshell has been divided into 10 crystallites having approximately the same volume and number of spins, as depicted in the scheme of Fig. 5. Every crystallite has a different uniaxial anisotropy direction $\hat{n}_i$ taken at random. As for the values of the anisotropy constants, we have taken $K_C = 4.7\times10^4$ erg/cm$^3$ (the value corresponding to bulk maghemite) and have evaluated $K_S = 0.1$-$1$ erg/cm$^2$ by considering the effective anisotropy obtained from the magnetization measurements as $K_{eff} = K_C + \frac{S}{V} K_S$ (being V and S the particle volume and surface, respectively). When expressed in units of K/spin, as used in the simulations, these values translate to $k_C \geq 0.01$ K and $k_C \geq 1$-5 K. Note that hollow polycrystalline particles, like the ones experimentally analyzed here, have 8950 spins, from which 91% are surface spins.

In Fig. 5, we display a snapshot of the low temperature magnetic configuration for $k_S = 30$ K attained after cooling from a disordered high temperature phase in zero applied magnetic field. The spins corresponding to each crystallite are colored differently and, inside each crystal, core spins have been distinguished with a lighter color tone.[28] Inspection of the displayed configuration shows that core spins tend to order ferrimagnetically along the local easy axes of each crystallite, while most of the surface spins remain in a quasi-disordered state induced by the competition between the surface anisotropy and AFM exchange interactions. The exchange interaction among the individual crystallites forming the shell is not sufficient to align all the magnetic moments of each crystallite in the same direction for the entire shell. That is, the magnetic behavior of hollow maghemite nanoparticles at low temperature is dominated by the crystallographic anisotropy of the individual crystal domains forming the shell.

In order to demonstrate the peculiar magnetic behavior of the nanoparticles associated to their hollow structure, we have simulated hysteresis loops for polycrystalline particles with different shell thicknesses; from a solid particle to a hollow particle with shell thickness similar to those of the particles experimentally characterized in this work. The hysteresis loops at low temperature (T= 0.5 K), were simulated by cycling the magnetic field between h= ±100 K in steps of 1 K. In figure 6, such hysteresis loops are shown for a particle with a fixed radius of 4.88 *a* and two values of the shell thickness $D_{Sh} = 1.92\ a$ (experimental hollow) and 4.88 *a* (filled). As compared to the loops of filled particles, the hysteresis loops of the hollow particles show increased coercivity, decreased remanence and remain open to larger fields with no saturation. This observation demonstrates that, despite having the same number of crystallographic domains, the hollow nanoparticles display distinct magnetic behavior with respect to filled particles. Results for decreasing values of the shell thickness indicate a progressive change in the magnetic response of the particles: As the shell thickness is decreased to the experimental value ($D_{Sh}$= 1.92 *a*), the increasing number of surface spins of the crystallites, together with their random anisotropy directions is responsible for the magnetic behavior of the nanoshells

The role of an increased surface anisotropy with respect to bulk for a hollow particle with the real dimensions can be understood by looking at the hysteresis loops computed for different values of $k_S$ shown in Fig. 7. When increasing surface anisotropy, the loops become more elongated, and they have lower high field susceptibility and higher closure fields. The qualitative shape of the loops for $k_S$> 10 K becomes similar to that of the measured ones shown in Fig. 4(a), demonstrating that the magnetization dynamics of real samples is dominated by the high proportion of spins on the outer regions of the crystallites forming the shell and their increased surface anisotropy. Moreover, by looking at the contribution of the core spins presented in panel (b) of Fig. 7, we see that the hysteresis loop of the core spins changes from square shaped to elongated with increasing $k_S$, indicating the increasing influence of the disordered surface spins on the reversal mode of the individual crystallites and of the whole hollow particle, which confirms the previous conclusion.

In Fig. 8, the simulated hysteresis loops obtained after field cooling the particle from a high temperature disordered state down to T= 0.5 K in different fields are shown. From these simulations, an appreciable shift of the hysteresis loop towards the left of the applied field axis can be observed for $k_S$= 30 K. Similar shifts were also experimentally obtained after field cooling the hollow particles. This loops shift is certainly due to the fact that, for high $k_S$ values, the applied field is not enough to saturate even the core spins. Therefore, the computed loop is a minor loop and the shift should not be erroneously ascribed to any exchange bias effects.

## Conclusions

At low temperature, non-interacting maghemite ($\gamma$-$Fe_2O_3$) hollow nanoparticles obtained by the Kirkendall effect show a ferrimagnetic-like behavior. However, their spins struggle to follow the external magnetic field, which results in low magnetic moments, high coercive and irreversibility fields and no magnetic saturation. This observation is associated to the particular arrangement of the crystallographic domains in the hollow geometry and to a high effective anisotropy, which arises from the extended amount of pinned spins at the surfaces and interfaces of such polycrystalline nanostructures (91% on 8 nm particles). The Monte Carlo simulations allow us to determine the role of the microstructural and geometric parameters on the magnetic behavior of hollow nanoparticles at the different temperature regimes. At low temperature, the exchange interactions between spins with different crystallographic easy axis inside the shell have a noticeable but not dominant influence on the hysteresis loops. The crystallographic anisotropy acts as glue fixing the spin orientation following the anisotropy axis of the randomly oriented crystallographic domains. In this scenario, the exchange interaction between different crystallographic domains inside thin polycrystalline shells is not sufficient to align the magnetic moment of each crystallite into a unique direction. As a result, the hysteresis loops resemble those of frustrated and disordered magnets such as random anisotropy systems. At high enough temperatures, thermal agitation permits spins of the different crystallite cores to detach from crystallographic anisotropy axis and to follow the applied magnetic field and the weaker intercrystal interactions. In this way, in the superparamagnetic regime, the spins of the crystallite cores within the shells tend to align coherently throughout the entire particle.

## Acknowledgements

This work was supported by the Director, Office of Science, Office of Basic Energy Sciences, Materials Sciences and Engineering Division, of the U.S. Department of Energy under Contract No. DE-AC02-05CH11231. A.C. thanks financial support from the Generalitat de Catalunya, Departament d'Universitats, Recerca i Societat de l'Informació. V. F. P. thanks financial support from MAT2006-13572-C02-02. Ll. B. thanks financial support from Spanish MCyT (MAT2006-13572-C02-01) and Consolider–Ingenio 2010 CSD2007-00041. O. I. and A. L. thank financial support from Spanish MCyT (MAT2006–03999, NAN2004-08805-CO4-01/02 projects) and Consolider–Ingenio 2010 CSD2006–00012. We acknowledge CESCA and CEPBA under coordination of C4 for computer facilities. We thank Prof. J. Long and his group for the assistance and use of their SQUID.

## Figure captions

**Fig. 1** Transmission electron microscopy micrographs of the hollow (a)-(c) maghemite nanoparticles. Scale bars correspond to 100 nm for (a) and 4 nm for (b) and (c).

**Fig. 2** (color online) ZFC-FC magnetization curve measured at 100 Oe. The red solid line corresponds to a fitting with a Curie law (M~1/T) of the experimental data in the SPM regime. The inset shows the blocking temperature of solid and hollow nanoparticles as a function of the average volume of material per particle (for hollow particles, the average volume of the cavity has been substracted from the average total particle volume).

**Fig. 3** (color online) Temperature dependence of the real $\chi'$ (solid symbols) and imaginary $\chi''$ (empty symbols) parts of the ac susceptibility measured at different frequencies (square: 1 Hz; circle: 10 Hz; triangle: 100 Hz; diamond: 1000 Hz) with an oscillating magnetic field amplitude of 4 Oe. The inset shows the fitting of the blocking temperature dependence on the characteristic relaxation time extracted from $\chi'$ curves. In this analysis, the point corresponding to the peak of the dc ZFC curve has also been included assuming a characteristic time window for that experiment of about 50 s. This point is distinctively marked as an empty circle.

**Fig. 4** (color online) (a) ZFC (filled symbols) and FC (10 kOe, open symbols) hysteresis loops at 5 K for the hollow nanoparticles. (b) Isothermal magnetization curve in the SPM regime measured at 200 K (empty circles) and fit to a distribution of Langevin functions plus a paramagnetic contribution (solid black line). The red dashed and blue dot-dashed lines show the contribution of the crystallite cores and surface spins to the fit. The inset shows the fitted distribution of magnetic moments of the ferrimagnetic component corresponding to spins at the crystallite cores.

**Fig. 5** (color) Low temperature snapshots of the magnetic configuration of a hollow particle with external radius R= 4.88 *a* and thickness $D_{Sh}$= 1.92 *a* with $k_S$=30 K as obtained from the Monte Carlo simulation. The upper (lower) panel shows a cut through a diametric plane parallel to the Z (XY) axis. The spins belonging to different crystallites have been distinguished with different colors, with core spins (those with bulk coordination) colored lighter.

**Fig. 6** (color online) Low temperature (T= 0.5 K) simulated hysteresis loops for a particle with $k_S$ = 30 K, external radius R= 4.88 *a* and two values of the shell thickness $D_{Sh}$ = 1.92 *a* (hollow particle), and $D_{Sh}$ =4.88 *a* (filled particle).

**Fig. 7** (color online) Low temperature (T= 0.5 K) simulated hysteresis loops for a hollow nanoparticle with external radius R= 4.88 *a* and shell thickness $D_{Sh}$= 1.92 *a* for different values of the surface anisotropy constant $k_S$= 0.01, 10, 30 K. Panel (a) shows the total magnetization, and panel (b) displays the contribution of the core spins only.

**Fig. 8** (color online) Simulated hysteresis loops for a particle with $k_S$ = 30, R= 4.88 *a* and $D_{Sh}$ = 1.92 *a* obtained after field cooling from a high temperature disordered state down to T= 0.5 K in different fields $h_{FC}$= 50 K (red cicles) and $h_{FC}$= 100 K (blue squares). The hysteresis loop obtained after cooling in zero field is shown in dashed lines.

**FIGURE 1**

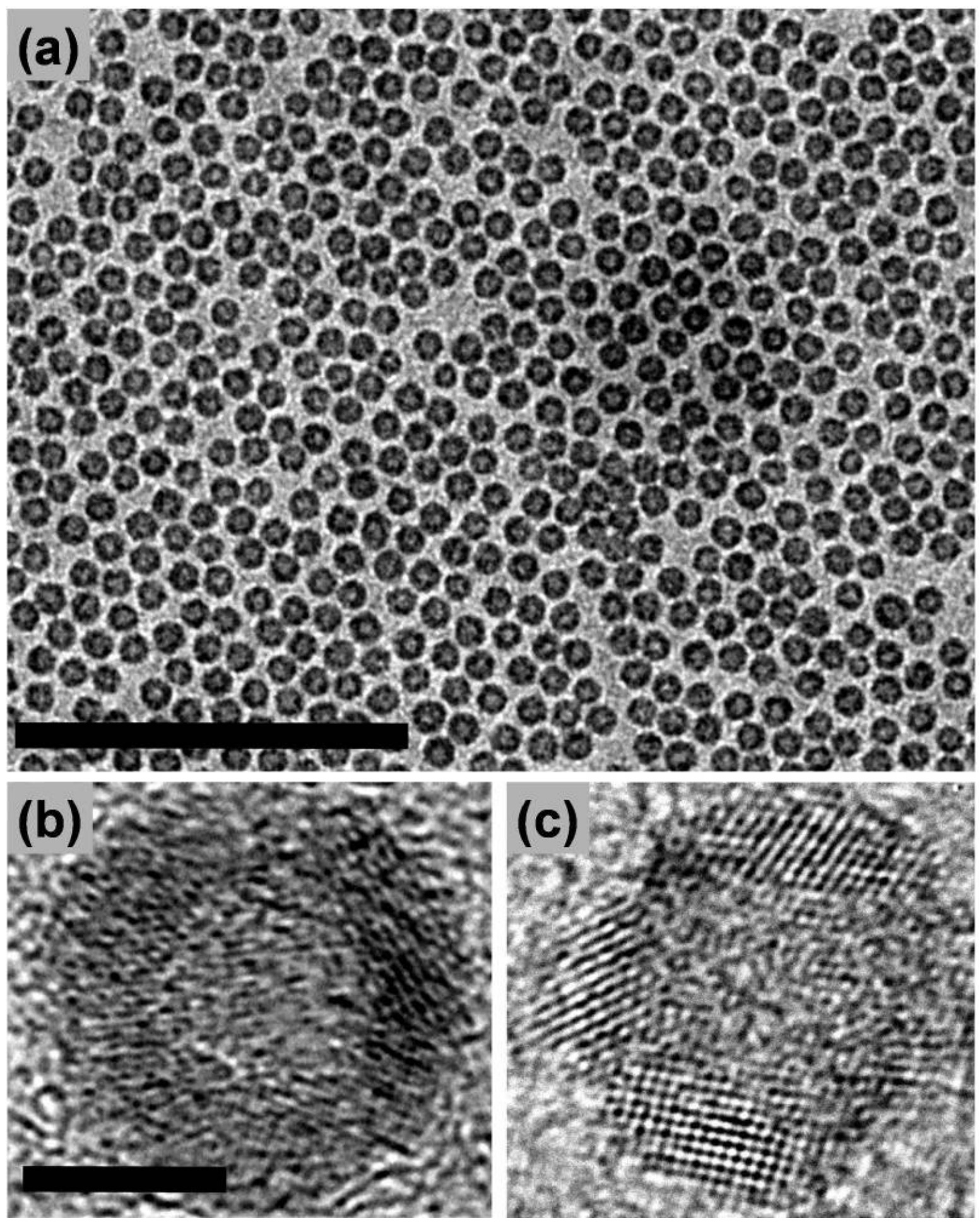

**FIGURE 2**

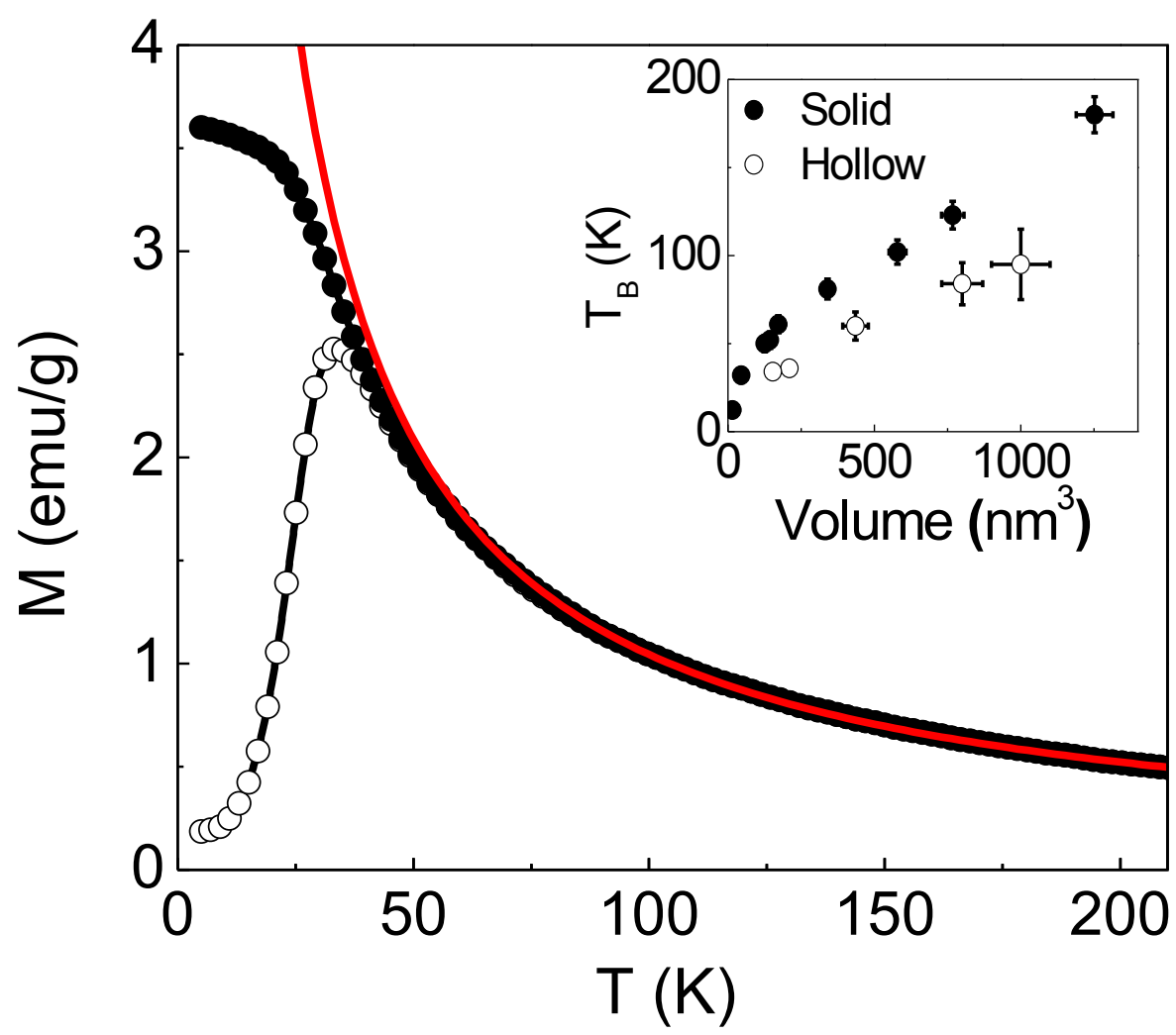

**FIGURE 3**

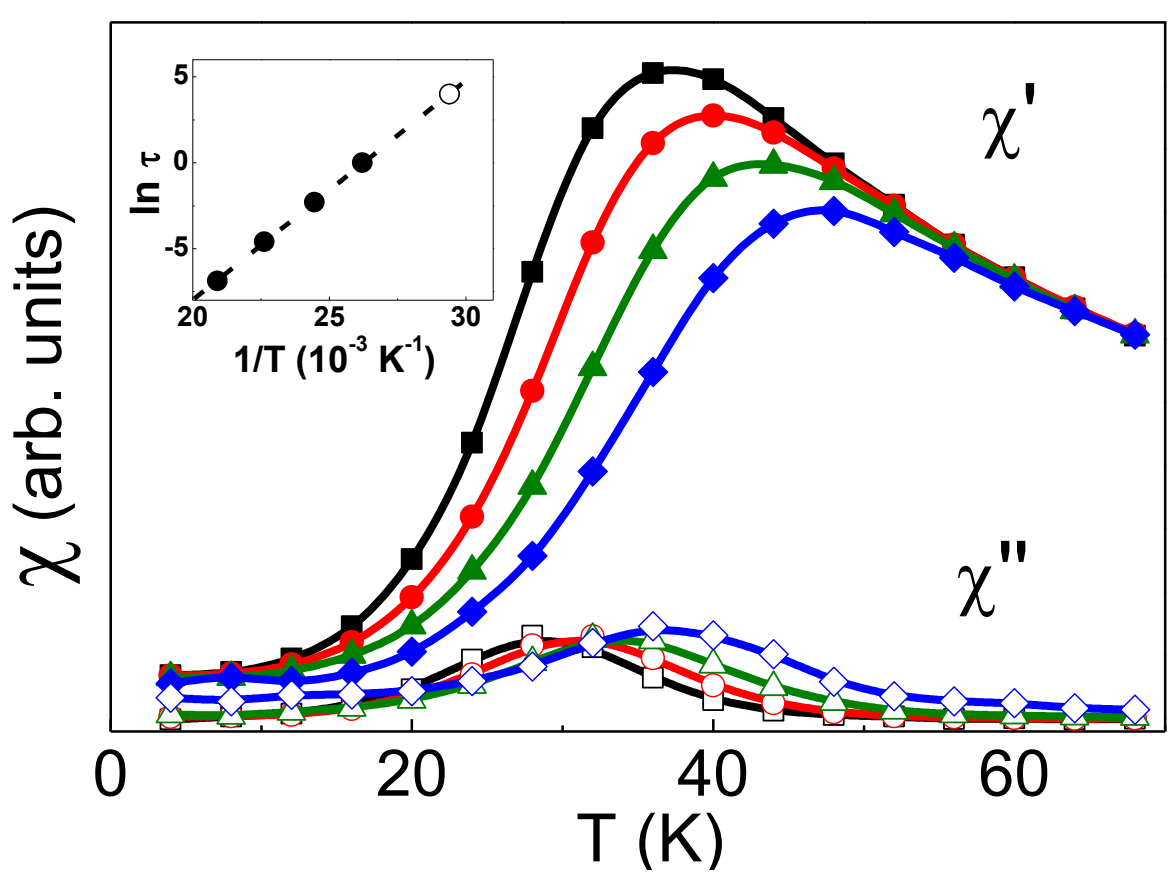

**FIGURE 4**

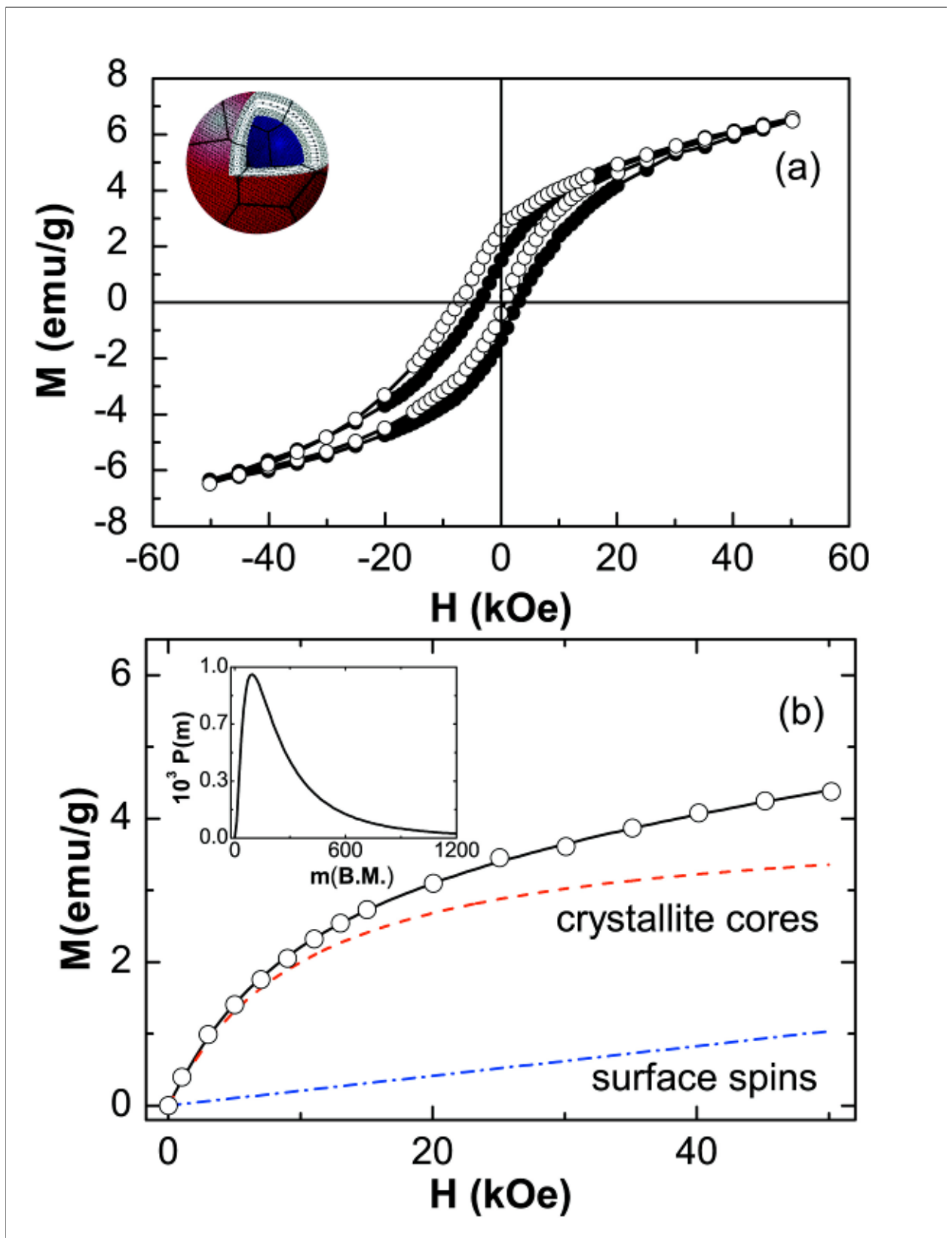

(a)
M (emu/g)
H (kOe)
(b)
M(emu/g)
H (kOe)
$10^3$ P(m)
m(B.M.)
crystallite cores
surface spins

**FIGURE 5**

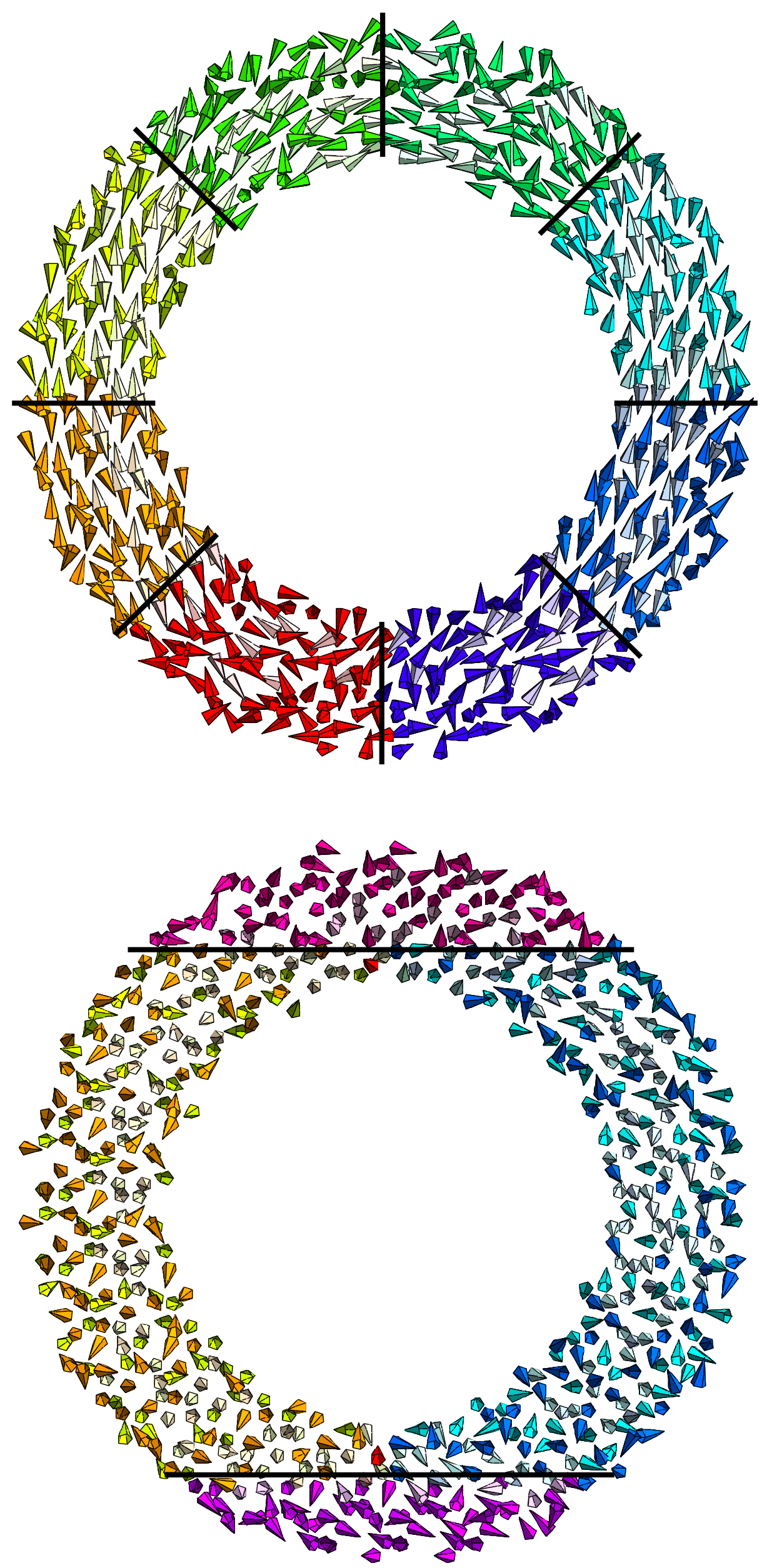

**FIGURE 6**

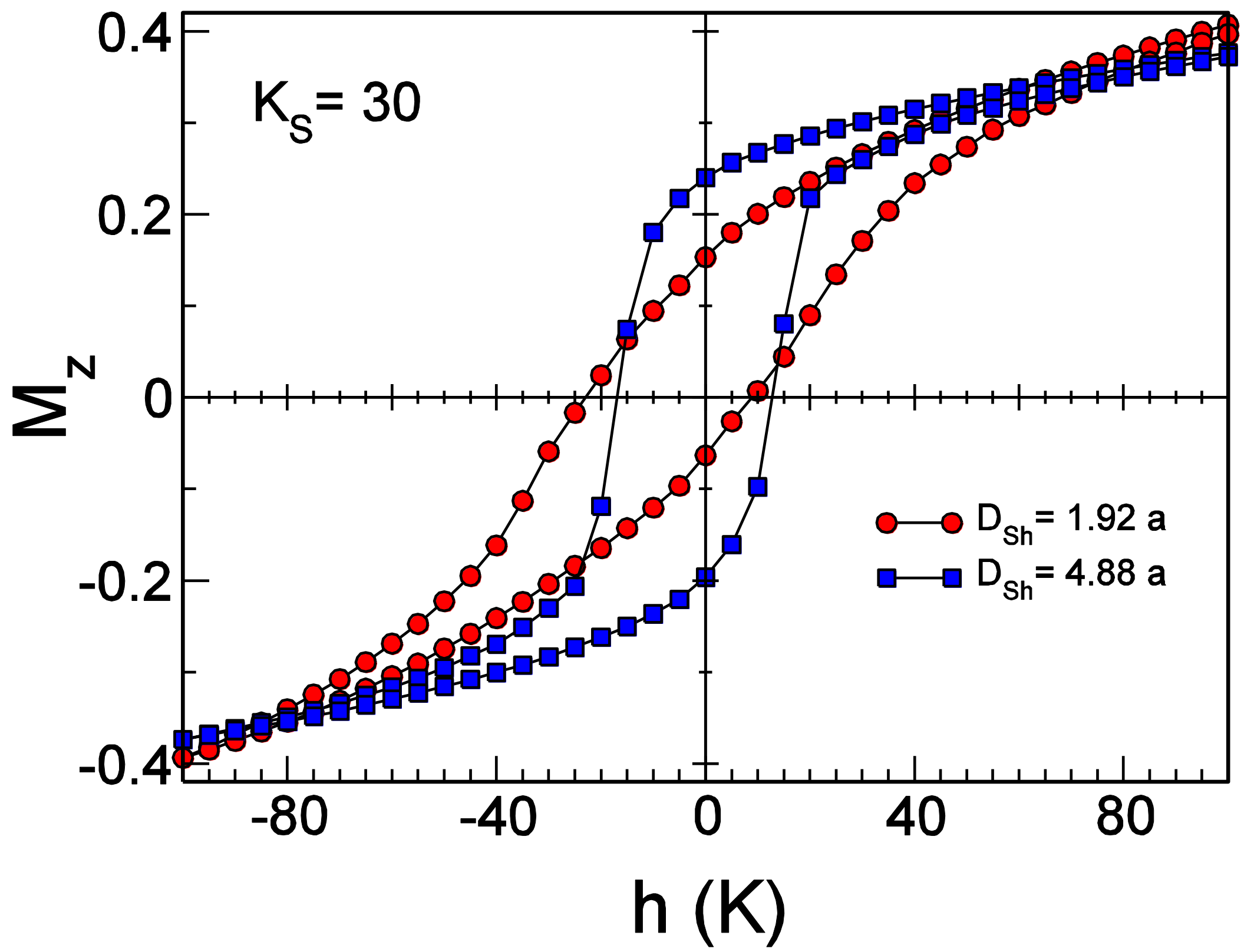

**FIGURE 7**

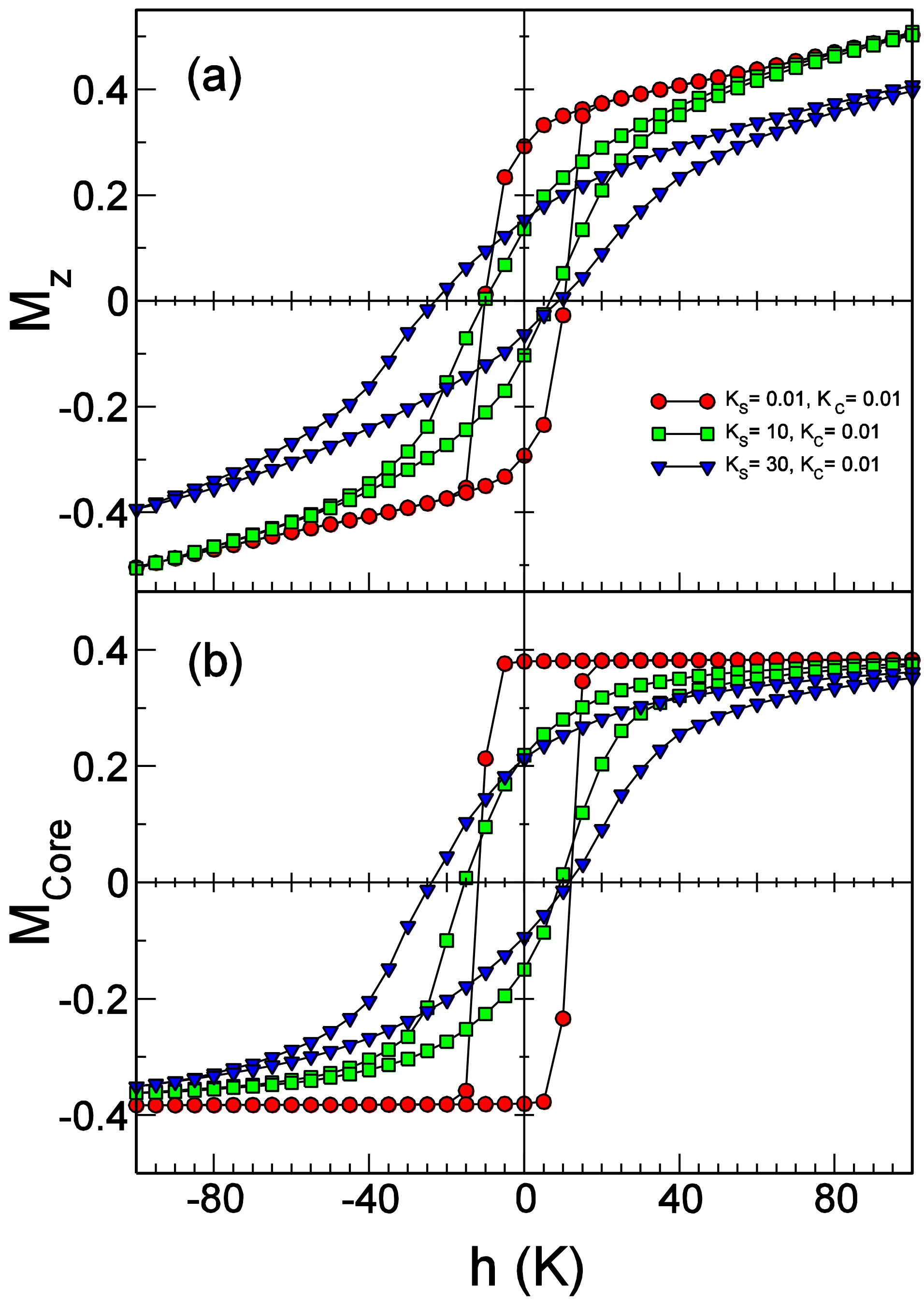

(a)
0.4
0.2
0
-0.2
-0.4
$M_z$
$K_S$= 0.01, $K_C$= 0.01
$K_S$= 10, $K_C$= 0.01
$K_S$= 30, $K_C$= 0.01
(b)
0.4
0.2
0
-0.2
-0.4
$M_{Core}$
-80
-40
0
40
80
h (K)

**FIGURE 8**

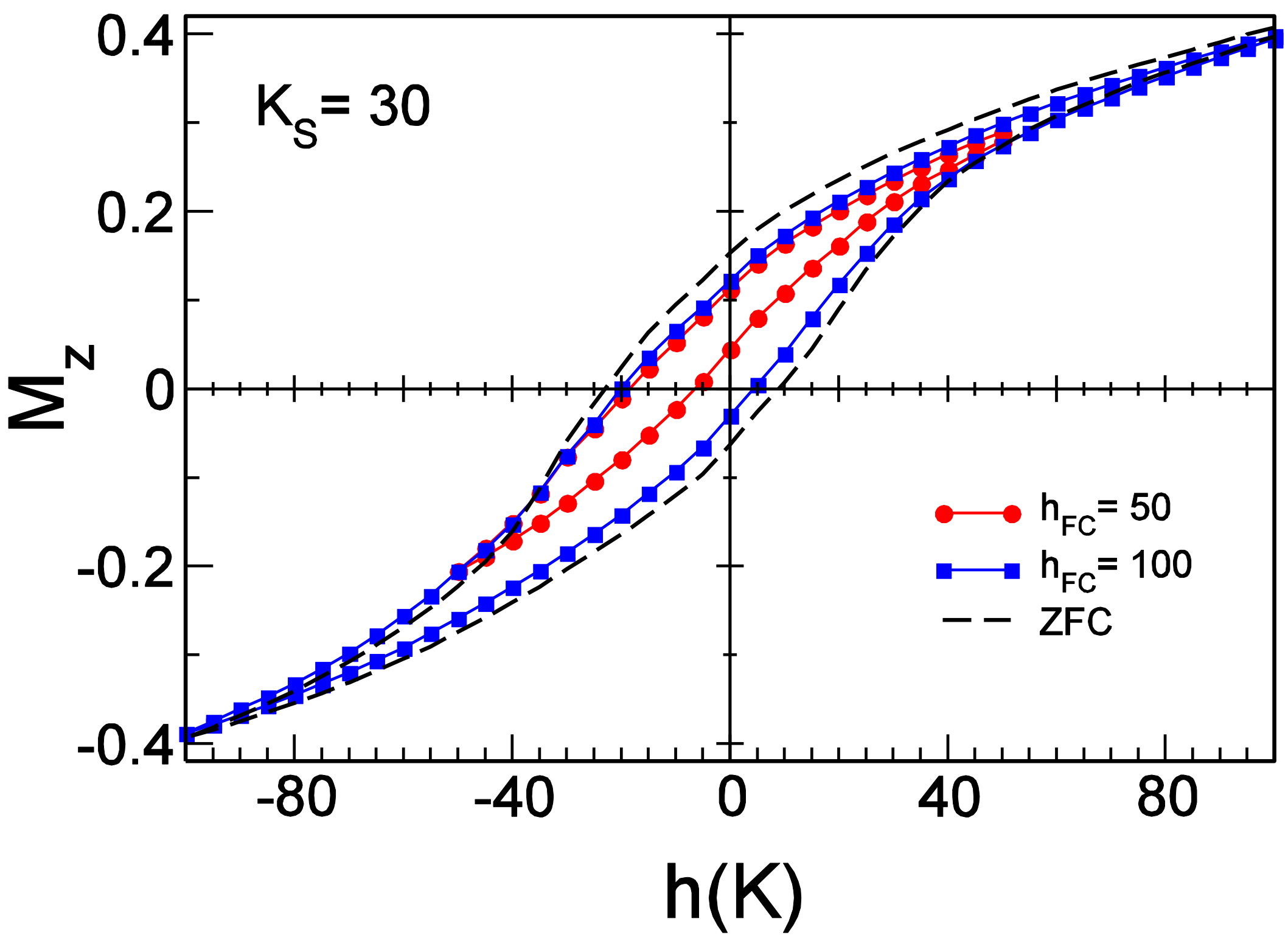